# Edge-computing Enabled Next Generation Wireless Networks: A Novel Approach to Provide Secure Multicast Services

Farid Sorouri
Department of Computer Engineering, Tehran Branch, Islamic Azad University, Tehran, Iran
Email: f.sorouri@iau-tnb.ac.ir

**ABSTRACT:**
Smart grids have received much attention in recent years in order to optimally manage the resources, transmission and consumption of electric power. In these grids, a comprehensive communication infrastructure is required to transmit the information required for observing, monitoring, making decisions, and applying load management controls. In these grids, one of the most important communication services is the multicast service. Providing multicast services in the smart communicative grid poses several challenges, including the heterogeneity of different communication media and the strict requirements of reliability, security and latency. Wireless technologies and PLC connections are the two most important media used in this grid, among which PLC connections are very unstable, which makes it difficult to provide reliability. In this research, the problem of geographically flooding of multicast data has been considered. First, this problem has been modeled as an optimization problem which is used as a reference model in evaluating the proposed approaches. Then, two MKMB and GCBT multicast tree formation algorithms have been developed based on geographical information according to the characteristics of smart grids. Comparison of these two approaches shows the advantages and disadvantages of forming a core-based tree compared to a source-based tree. Evaluation of these approaches shows a relative improvement in tree cost and the amount of end-to-end delay compared to basic algorithms. In the second part, providing security and reliability in data transmission has been considered. Both Hybrid and Multiple algorithms have been developed based on the idea of multiple transmission tree. In the Hybrid algorithm, the aim is to provide higher security and reliability, but in the Multiple algorithms, minimization of message transmission delay is targeted. In the section of behavior evaluation, these two algorithms have been studied in different working conditions, which indicates the achievement of the desired goals.

## 1. INTRODUCTION

Nowadays, one of the most important issues in smart grids we face is heterogeneity. This means that we are faced with different communication media in different situations of the smart grids. Also, different smart media are usually used to transmit data, depending on whether we are talking about HAN, NAN or WAN. At the WAN level, various network technologies such as optical fiber, leased lines and WiMAX are used. Technologies such as optical fiber, WiMAX, PLC, satellite communication and cellular networks are used for distribution networks that connect NANs (Neighbor Area Network). At the radio access network level, short-range wireless technologies such as Bluetooth, ZigBee and WiFi are used.

Therefore, there is a need for connections among heterogeneous networks. This heterogeneity leads to problems in providing multicast services. For example, the crying baby problem is caused by the heterogeneity in this network. This means that the transmitter in the multicast group must coordinate itself with the receiver with the lowest received bandwidth. This issue reduces network throughput. On the other hand, due to the heterogeneity, one of the most important media used in this network is PLC connections. In [1], many problems that these links are encountered for data transfer was listed. Geocast, on the other hand, is one of the subclasses of the multicast problem. In Geocast, the point that makes different nodes a member of a multicast group is their geographical location and their placement within a certain boundary. Geocast can have many applications in the smart grid.

If it is going to determine a problem be solved in this research, consider this scenario: control center wants to reduce consumption by announcing a high cost per kilowatt hour of consumption to the industrial parts of the city in order to meet peak consumption hours in the home sector. The control center should multicast a message containing the new cost for the next hours to the city's industrial area. In this study, we hypothesized that the control center should first transmit multicast packets to the wireless nodes located in the destination area via the urban WiMAX network, and then these wireless nodes deliver destination area of the packet into the nodes located in this area through the PLC connections. In fact, by considering the two wireless communication media and PLC, we have considered the heterogeneity of the smart grid. Delivery of these packets must meet the latency and reliability requirements which is in the standards defined for this protocol. In the meantime, the issue of security and reliability of network nodes is also raised.

Many protocols, such as DVMRP [2], CBT [3], and PIM [4] were introduced to provide multicast services on the Internet. Given the large volume of connections in the network, the need for multicasting arises by considering the heterogeneity of links. The RMX protocol [5, 6] or TOMA procedure [7] is among the works that has done in this field [8, 9]. Generally, the next generation wireless networks require impressive radio resource management (RRM) [10-15] and effective routing strategies for optimal power utilization [16-18], load balancing [19, 20], and security assurance [21, 22]. After the MANET networks were created and research work went toward these networks, some work was done to construct a multicast tree in these networks [23, 24]. In particular, the use of Geocast was considered in these networks, for which specific protocols were provided the same as [25, 26]. For smart grids, little effort has been made to provide multicast service on

PLC connections. For example, in [27, 28], a method has been only presented for effective multicast on PLC connections, but the issue of locating PLC connections along wireless connections in multicast has not been considered so far.

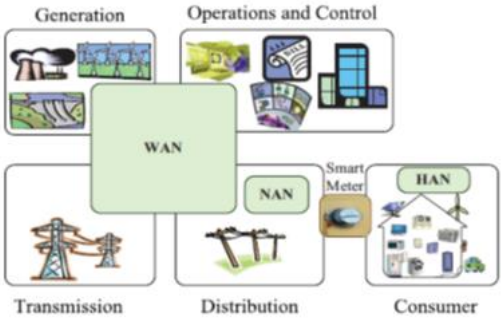

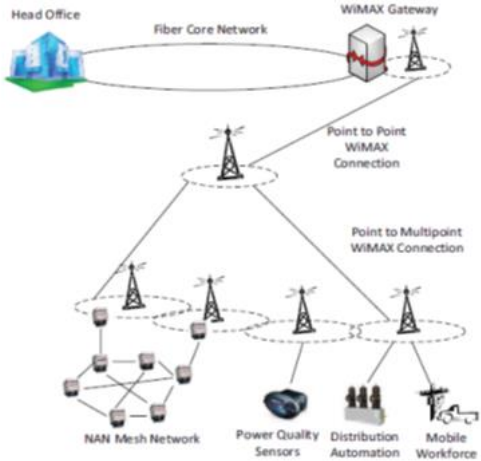

**Fig. 1**. The structure of the smart grid network

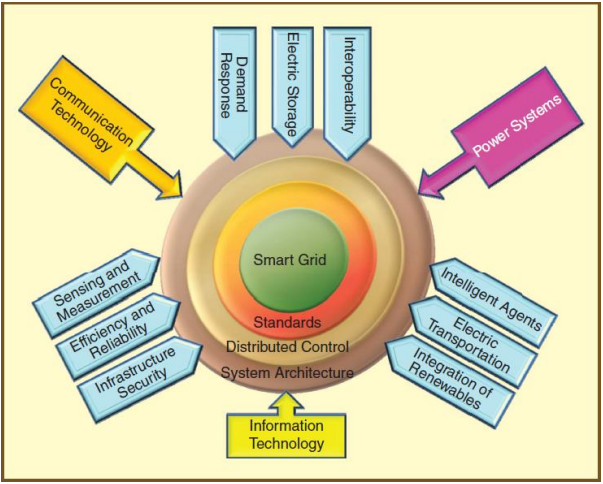

**Fig. 2**. The main characteristics of the smart grid system

The RMTP protocol [29] is a protocol based on a hierarchical structure and local grouping. Each group has a specific recipient who is responsible for the retransmitting. RMTP is a grouped DER protocol. SRM [30] allows all network nodes to retransmit.

Developments for local error retrieval are provided in this method. SRM therefore falls into the category of ungrouped DER protocols in this subdivision, and the developments provided for local retrieval in this algorithm fall into the category of grouped DER. The NP [31] algorithm can only retransmit multicast resources. Therefore, this algorithm falls into the CER category. The MESH protocol [32] is also known as a DER protocol that benefits from both local and global retrieval.

In this paper, two series of algorithms were presented to provide multicast security service in heterogeneous smart grids. The first series were algorithms worked with a single-step topology. That is, each PLC node had a wireless node next to itself with one step. MKMB and GCBT algorithms were proposed for this case. MKMB algorithm was an extension of KMB algorithm. The idea of GCBT method was to use geographical information in the CBT algorithm. An optimization model for constructing the Steiner tree was proposed and solved. The aim was to compare these algorithms with the optimal mode in terms of construction costs. These algorithms are compared with each other and evaluated for their performance. The result was that the MKMB algorithm is not much different from the usual KMB method in terms of performance, and the GCBT method tried to use geographic information alongside the CBT protocol, which its performance was good, but because it did not take into account the variable status of PLC connections, it cannot be implemented in smart grids.

Also, two hybrid and multiple algorithms were presented. Their aim was to provide reliability for the nodes in the destination area by transmitting two wireless nodes into the destination area. we compared the two algorithms in terms of tree cost and average reliability of nodes in different scenarios and interpreted the obtained results. The main achievements of this research are:

1. To provide reliability to ensure that packets are received by nodes of the destination area by creating two optimal trees to transmit packets to each node.
2. To consider the security issues by selecting the wireless nodes with the most reliability.
3. 100% coverage of nodes in the destination area in terms of receiving multicast packets.
4. To reduce delay for arriving packets to destination nodes, which is useful for delay-sensitive applications in smart grids.
5. To provide a way to incorporate geographic information into the KMB algorithm in order to use this additional information for better routing in packet multicast process.
6. To incorporate geographic information into the CBT algorithm and provide GCBT method.

## 2. SYSTEM MODEL AND PROBLEM FORMULATION

As mentioned earlier, the Steiner Tree problem is an NP-Complete problem. Therefore, to construct this tree, Integer Linear Programming (ILP) was used [33]. In the following, written model is introduced and its different parts are discussed.

Minimize $\sum_{i \in V} \sum_{j \in V} Y_{i,j} * W_{i,j}$ over : X, Y
$subject\ to$ :

$$Y_{i,j} \leq E_{i,j} \quad \forall\ i, j \in V \quad (1)$$

$$X_{i,j,k} \leq Y_{i,j} \quad \forall\ i, j \in V\ \text{and}\ k \in D \quad (2)$$

$$X_{j,i,k} \leq Y_{i,j} \quad \forall\ i, j \in V\ \text{and}\ k \in D \quad (3)$$

$$\sum_{<s,v> \in dplus\ (src)} X_{s,v,k} - \sum_{<v,s> \in dminus\ (src)} X_{v,s,k} = 1 \quad \forall\ k \in D \quad (4)$$

$$\sum_{<s,v> \in dplus\ (D[k])} X_{s,v,k} - \sum_{<v,s> \in dminus\ (D[k])} X_{v,s,k} = -1 \quad \forall\ k \in D \quad (5)$$

$$\sum_{<v,s> \in dplus\ (v)} X_{v,s,k} - \sum_{<s,v> \in dminus\ (v)} X_{s,v,k} = 0$$
$$\forall\ k \in D, \forall\ v \in V - \{src\ \cup\ k\} \quad (6)$$

In this model, we have a flat network consisting of N nodes. For simplicity, it is assumed that all nodes have the same features, and in particular have the same radio range equal to R. The set of nodes is V= $\{v_0, v_1, \ldots v_n\}$. In the following, the node $i$ and the position of node $i$ will be shown with $v_i$. In addition, the position of each node is determined by a pair of coordinates $v_i = (x_i, y_i)$. Suppose that $d(v', v'')$ is the Euclidean distance between the positions $v' = (x', y')$ and $v'' = (x'', y'')$. Cumulative distance is defined as the distance between node v and destination nodes $V^D$, where $V^D \subseteq V$, and this distance is donated by cd $(v, V^D)$. In fact, the cumulative distance is the sum of the distances between position v and the position of each node in $V^D$.

$$cd\ (v, V^D) = \sum_{\forall w\ \in V^D} d(v, w) \quad (7)$$

Another concept that has been introduced is cumulative progress. Consider two points: $v'$ and $v''$. The cumulative progress of node $v'$ to $v''$ towards nodes $V^D$, which is denoted by cp $(v', v'', V^D)$, is considered as the difference of the cumulative distance of node $v''$ from the cumulative distance $v'$.

$$cp(v', v'', V^D) = cd\ (v', V^D) - cd\ (v'', V^D) \quad (8)$$

This cumulative progress criterion can now be used to compare different options for selecting the next relay node.

In this method, first the link with the most cumulative progress in the network is selected among the existed links. Then the cp value of each link is converted by the following formula.

NewCP=-1*cp + max           (9)

The lower value of NewCP is desirable. For example, if cp = max, the NewCP = 0, which is the lowest value. This formula also covers negative cp values. Negative values are related to the return edges. For example, consider the figure below.

Then the final weight of each edge is determined by the function f as follows:

$$F(v', v'') = A * NewCP(v', v'', V^D) + B * Delay(v', v'') \qquad (10)$$

Depending on the importance of each factor of the distance to the destination area and the delay are assigned to the values related to A and B. Of course, it must be considered that the delay and distance values must be in the same range in order to see the effect of each. To do this, numbers related to delay are scaled to the interval numbers.

It was mentioned earlier that the first step of the proposed KMB method requires the implementation of the shortest path algorithm between each pair of nodes in the set S. The shortest path algorithm used to implement is the A* method. Before performing this step, the weight of the links is calculated according to the function F.

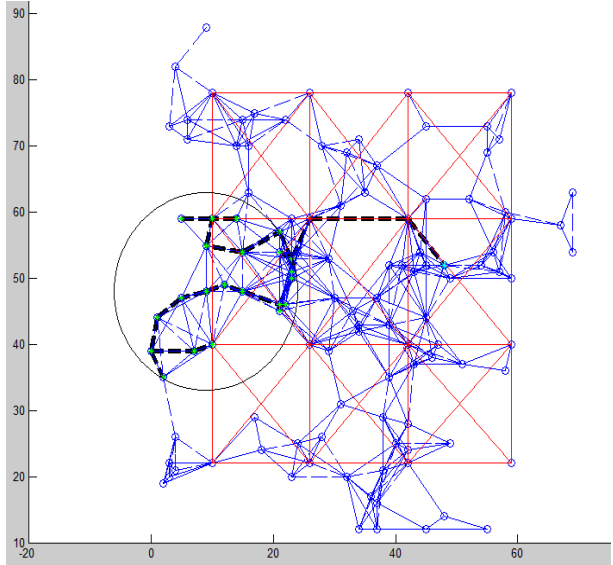

**Fig. 3**. The resulted tree of execution the MKMB algorithm is indicated with the black color

### 3. THE PROPOSED MULTIHOP ALGORITHM

The algorithms to be talked about in the previous section focused on reducing the cost of the tree. In this section, we want to present two other algorithms that are more complete than the previous algorithms and also include the security and reliability factor.

In multicast articles, IP refers to a set of four interrelated problems. These four problems are:

1) Routing: Finding a route from one transmitter to other members in a multicast group
2) Reservation: To serve resources along this route
3) Reliability: To ensure that despite the loss of packets and their failure, all packets reach from one transmitter to all receivers
4) Flow control: To adjust the rate at which a transmitter transmits packets in a way that intermediate links and receivers are not overloaded. These problems are more or less perpendicular to each other [34].

The protocol we are going to introduce in this section focuses on sections 1 and 3 of above, that is routing and reliability in multicast, because one of the most important requirements of the smart grid is the issue of reliability. This means that by creating backup routes and additional links, something must be done that if a link faces an error or congestion occurs, the package can be transmitted from an alternative route. Of course, this issue contradicts the delay factor while creating a tree. This is because creating a tree with more branches and non-optimal alternative paths can help increase reliability, but this increases the transmission delay. Of course, the issue of reliability is raised for the graph, not the tree, because in the tree, we have only one path to reach each node, and if we want to have two paths, then it will no longer be a tree and we will have a graph. Therefore, in order to have both delays by forming an optimal tree and increase reliability by creating separate paths, we came up with the idea that the nodes in the destination area are members of two transmission trees. These two trees are created by two wireless nodes connected to the destination area. We will talk more about this idea later.

The optimization model used to determine the location of wireless nodes is as follows:

$$\text{Minimize } \sum_{r=1}^{R} Z_r$$

Over: Z, X
Subject to:

1. $X_{n,r} \leq Z_r * G_k(n,r) \qquad \forall n, \forall r \qquad (11)$
2. $\sum_{r=1}^{R} X_{n,r} = 1 \qquad \forall n \qquad (12)$

An important point to be considered is the level of awareness of the control center about this topology. We assume that the control center is aware of the position of the nodes relative to each other and the topology as a whole because the topology is static. Now the question may arise as to why the control center does not construct the optimal tree itself with its comprehensive knowledge of the network.

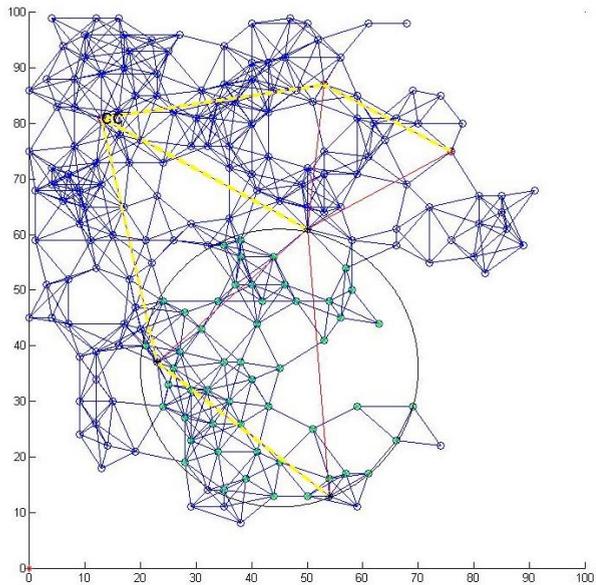

Fig. 4. The applied multihop topology in the scenario with indicating the target areas and candidate nodes

The answer is that the changing conditions of PLC connections have made them so unreliable that they need to be constantly updated on their status in decision centers. In fact, by transmitting these routing packets, the wireless nodes want to be aware of the latest status of PLC connections so that they can construct two trees based on the protocol intention.

The steps for doing this protocol, the exchanged packets and the function of each node are given below:

1. The control center wants to transmit a multicast message to a specific geographical area marked with a circle. The first thing it does is to determine the nodes in the destination area. Assume that c is the number of nodes in the destination area. The control center then creates an array of length c that corresponds to the id of the destination nodes. Each cell of the id array stores a node of the destination area. We call this array an area. It then identifies candidate nodes. Candidate nodes are nodes that are connected to at least one node (PLC or wireless) in the destination area. The nodes marked in black color in Figure 4 are the candidate nodes.

2. The control center then transmits the multicast packet to the candidate nodes through the tree. This transmission can either be the same as methods based on core to all wireless nodes (all of which are core) or it can only transmit to candidate nodes through source routing. This initial packet, in addition to containing the multicast data, includes the center point of the circle of the destination area and its radius, as well as the area array.

All candidate nodes receive the initial packet from the control center. Each then broadcasts a routing packet through the PLC nodes from the destination area, to which they are connected, to the destination area. The contents of this packet consist of four sections:

The center of the circle of the destination area and its radius is taken from the control center Wireless node itself. This id is needed because all candidate nodes in this packet transmit all broadcasts to the destination area and the PLC node needs to distinguish between these packets. An auxiliary array called parent whose length is c and is used to form a tree. In fact, for example, if parent [2] = 6, it means that the father of node 2 is node 6 in this tree. The initial value of this array is one of the following three conditions based on the status of node i: If i is the id node, it means that it is the node of a candidate that is in the destination area itself. Initial value -2 is assigned to nodes If the candidate node and the id identifier are outside the destination area, the initial value -2 is assigned to nodes that are one step away from it and are located in the destination area. Otherwise, the initial value is -1

### 3.1. The Proposed Hybrid Algorithm

In this method, the goal is the security and reliability of wireless nodes. That is, we prefer reliability over delay and reliability to nodes. The various steps of the algorithm are presented as the following:

There are three points in this algorithm that need further explanation:

1) The candidate node, located in the destination area, itself becomes the first person responsible for transmitting to itself. In the continuation of the implementation process of the program, the second responsible person will be determined.

2) This is determined by examining the parents array sent by the wireless nodes. If, for example, parents [i] related to candidate id has a value of -1 according to the initial value of parents array, it is clear that id cannot have 100% coverage.

3) At this point in the flow graph, it may not be practically possible count = 1, that is there is only one node that is eligible for 100% coverage. Because if there is a node that has 100% coverage, and there is also another candidate node, it would certainly be selected and make count 2. Because that node is also connected to one node in the destination area, and it can access all of the destination area through it. Therefore, it is impossible to find count = 1, that is there is only one node with 100% coverage. The value of the count is either 0 or 2 or more. On the other hand, because of the issue of node reliability, we cannot reach a suitable responsible assignment. We have to use more than 2 candidate nodes to transmit in order to work like a case where multiple protocols work. Therefore, we use the multiple algorithm in this section to both cover 100% and have trees with the minimum possible depth. This is where we come to the meaning of the word "possible" in defining the purpose of the Hybrid algorithm. These are the problems that make it impossible to find 2 candidate nodes through which we can make all the nodes in the destination area a member of two separate trees.

In the following, an example of the result of implementation on topology [35] is given. The result of the implementation and the resulting trees are as follows. In fact, in this example, there are 4 candidate nodes including nodes 45, 47, 44 and 48. Among these 4 nodes, nodes 44 and 45 have been selected according to the above algorithm. A tree shown with yellow links has node 45 as its root, and the tree marked in red has node 44 as its root. As you can see, each node in the destination area is a member of two trees, and in general, two candidate nodes have been assigned to construct the tree.

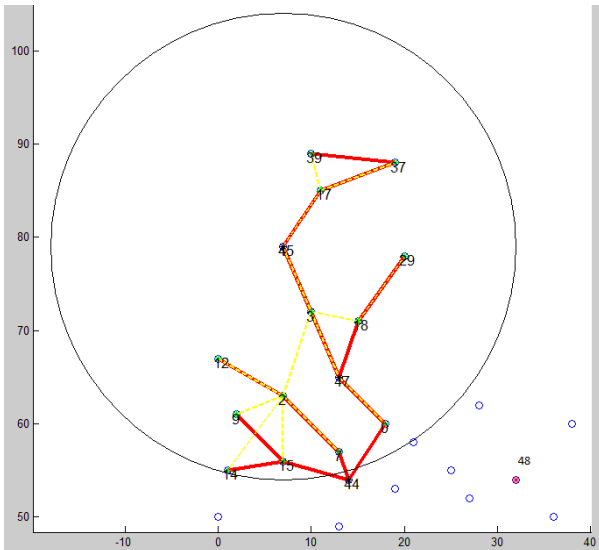

**Fig. 4**. A sample of the achieved results by the proposed hybrid method

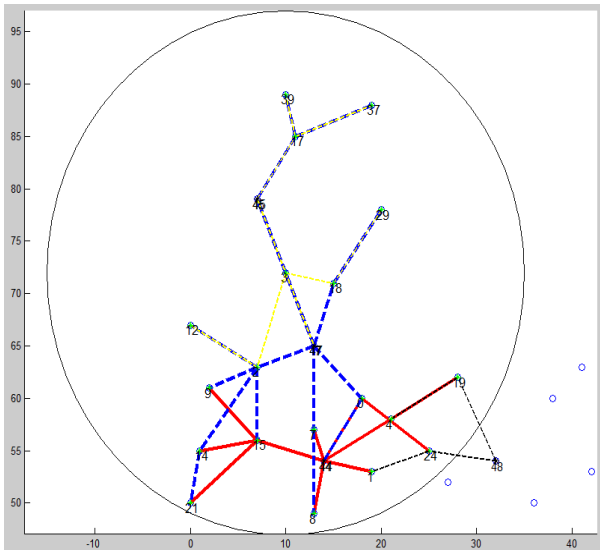

**Fig. 4**. A sample of the achieved results by the proposed multihop method

## 4. NUMERICAL AND SIMULATION RESULTS

In this section, we evaluate the proposed algorithms through simulation. Implementation of all proposed algorithms with C ++ programming language. The C code was typed in the Linux environment and was compiled by the GCC compiler version 4.4.3 of the programs. During the implementation of the programs, any ready-made library related to the network was not used. We created the required topology ourselves, which in previous sections, we created these topologies according to the structure of the smart grid. SCIP solver was also used for linear programming models. The language of communication with this solver is zimpl, the language in which models were written and given to solve the SCIP program. This solver was used in the Linux environment. Linux scripts were used to perform repeated simulations with different values to speed up and facilitate the work. MATLAB software was used to display the status of the network and the generated trees. The failure related to the network topology seen in the previous sections is the output of this program.

The evaluation section of the proposed methods, like the section of the proposed methods, is divided into two parts: single-step and multi-step algorithms. The criteria used in each section are somewhat different, because the two have different requirements.

### 4.1. Evaluation of single-step algorithms

First, we want to examine the criteria by which we compared different algorithms. These criteria are:
 1. Tree cost: It means the total weight of all tree links formed.
 2. End-to-end delay (one-way delay): It means the time taken for the packet to be transmitted from the source to the farthest destination in the multicast tree. In this study, we have excluded delivery delay, queuing delay, and packet scheduling versus link propagation delay. Because propagation delay usually has larger values than other delays, we have only considered this delay. This delay also indicates the heterogeneity of the smart grid.

On the other hand, we have several parameters in our program. By changing these parameters, we calculate the mentioned measurement criteria. In fact, these parameters are different perspectives on a program. We want to examine the results of generated algorithms under different conditions. These parameters are:
 1. Radius of the destination area: Increasing the radius of the destination area leads to an increase in the number of nodes located in this area. It also increases both tree cost criteria and the end-to-end delay for different algorithms. The purpose of studying this parameter is to examine how to increase this criterion for different algorithms.

2. Different topologies: Because the generated topology is based on the generation of random numbers, thus different topologies can be generated by changing the value of SEED related to the functions of generating random numbers.

Now we examine the result of changing these parameters on different algorithms.

### 4.2. The parameter of the cost of the destination area

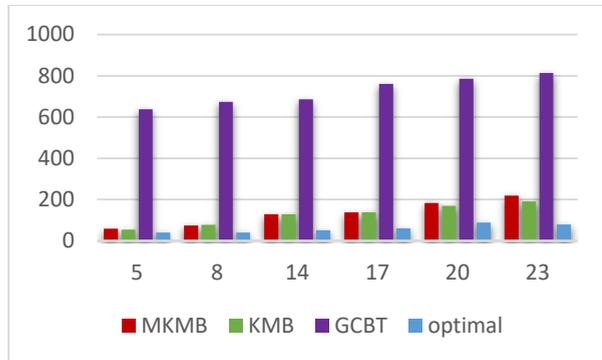

**Fig. 5.** The cost of the four optimized algorithms with respect to the various radius

Regarding the evaluation of KMB and MKMB methods achieved in figure 5, it should be said that to calculate the cost of the tree, we have compared the cost of trees obtained from the two methods with the assumption of the same link weight. In fact, if we calculated the weight of the MKMB tree based on the weight of the links including NewCP (section 1), the cost of the MKMB tree would definitely be more than KMB in all implementations. Thus, we stored the initial weights of the links that showed the delay in the MKMB algorithm. Then, the weight of the links was changed and the KMB algorithm was implemented. Finally, the weight of the resulted tree was calculated according to the initial weights of the links.

The reason the GCBT method costs is too distant from the rest methods is that it uses a shared tree. As can be seen, the tree cost values for the optimal state obtained by implementing the optimization model are the minimum values for different radiuses. The important point in this graph is the almost identical values of the KMB and MKMB algorithms. It shows that because the two algorithms both follow the same procedure, even with the weight of different links, the results are almost the same.

### 4.3. End-to-end delay

As figure 6 exhibited, it shows the end-to-end delay for the first part algorithms. Like the tree cost parameter, the GCBT method has the highest end-to-end delay because the source and destination nodes reach each other through a fixed tree, which in most cases, it is too distant from the optimal tree. Another important point is the values of the optimization model, which is indicated by the name "optimal" in the graph. The "optimal" name in the previous graph was a good name, because it provided the optimal tree cost and was lower than other methods, but in the end-to-end delay parameter, this algorithm is not optimal. However, there are models, while minimizing the cost of the tree, are bounded by a certain end-to-end delay - although they do not minimize it. The best values are related to the MKMB algorithm, which has the least end-to-end delay.

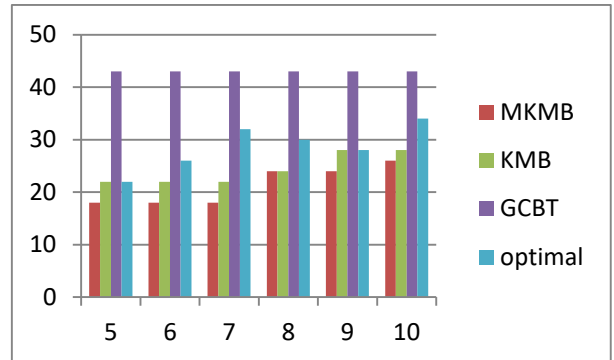

**Fig. 6**. The end-to-end delay of the four optimized algorithms with respect to the various radius

### 4.4. Evaluation of multi-step algorithms

In this section, we evaluate the performance of two Hybrid and Multiple algorithms. First, the criteria that are compared based on these two algorithms are given, then the parameters by which these criteria are measured are discussed. We compare these two algorithms in terms of two criteria: tree cost and average reliability of nodes. The parameters that we change are the radius of the destination area, the maximum number of steps that the PLC node reaches to a wireless node, and the size of the entire network. We have three parameters and two criteria that give a total of six different states and graphs.

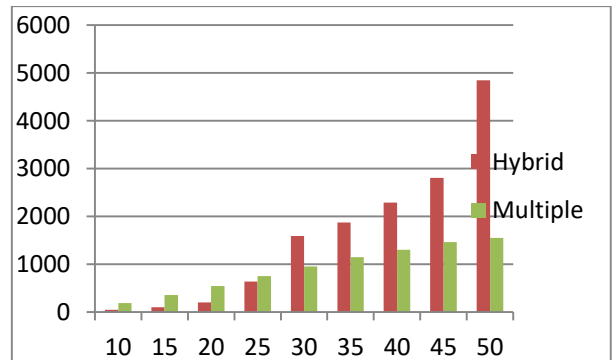

**Fig. 7**. The comparison of the Hybrid and Multihop approaches with respect to the various radius

## 4.5. The parameter of the radius of destination area

Based on Figure 8, the reliability of the trees of the two algorithms is compared when changing the radius of the destination area. In the hybrid algorithm, only two candidate nodes are assigned to construct the tree. Among all candidate nodes, only these two nodes receive the multicast packet from the control center. That is, of all the shared tree connections on wireless nodes, only the connections leading in these two nodes from the control center will be involved in calculating the cost of the tree. In the Multiple algorithm, however, all candidate nodes must receive a multicast packet, and as a result, all the wireless connections leading to them will be included in the tree cost. But on the other hand, the trees generated by the Hybrid algorithm inside the destination area are deeper because we want to involve only two nodes. But in the Multiple method, the trees generated in the destination area have less depth because all the nodes try to generate the tree. In fact, in the Hybrid algorithm, the cost of the tree related to the wireless part of the connection and the shared tree is low, and in the part related to the tree inside the destination area is high. Conversely, in the Multiple algorithm, the major cost of the tree goes back to the shared tree.

In fact, as the radius of the destination area increases, the cost of trees generated within the destination area will be increased, but the cost of the tree associated with the shared tree section on wireless connections remains almost constant because the radius must increase so much that the candidate nodes add a new set, so that links leading to this new node to be included in the tree cost calculations. Therefore, increasing the slope of the wireless part of the tree costs is slight and has a steep slope in the part of the destination area and PLC nodes. Therefore, approximately up to 25 radiuses, the total tree cost of the Multiple method is greater than the Hybrid method. Because the small destination area up to this radius is the total cost of the tree depending on the wireless part of the tree, in which the Multiple algorithm has a more branched tree. But with the increase of the radius, the opposite is the case. The cost part of PLC connections is higher than the part of wireless connections and the total cost of the Hybrid tree increases compared to the Multiple method.

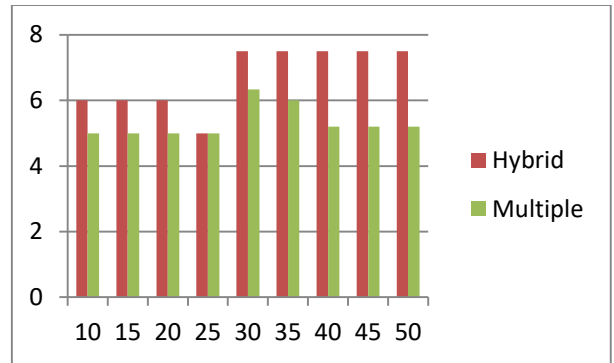

**Fig. 8**. The average of reliability for the proposed Hybrid and Multihop algorithms

## 4.6. Parameter of network size

The following graph shows the average reliability of the nodes involved in the multicast tree per radius of different destination areas. Given that the Hybrid method has a higher priority for selecting candidate nodes with higher reliability, it was expected that it would have higher average reliability than the Multiple method. But the interesting point about this graph is that as the radius of the destination area increases, the average reliability for the Multiple method remains almost constant and increases for the Hybrid method. The reason is that as the radius of the destination area increases, new candidate nodes enter the algorithm, and the Hybrid algorithm has more options for the two intended candidate nodes, thus its average reliability increases. But, in the Multiple algorithm, because all candidate nodes are involved in calculating the average reliability, by the entering of new candidate nodes, its average reliability not only does not improve considerably, but may also decrease as in the graph.

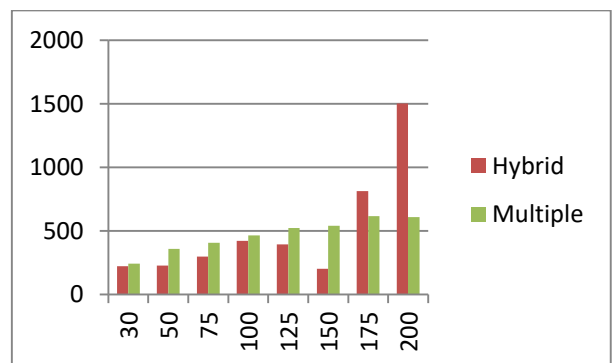

**Fig. 9.** The cost of the proposed Hybrid and Multihop algorithms with respect to the various network scales

The proposed multihop method provides different tree cost values for two algorithms for different network sizes. For this scenario, the maximum step size between the PLC and wireless nodes is 2 and the radius of the destination area is 20. This graph actually confirms the same results obtained from this figure. Of course, the reason why the cost of the tree for the Hybrid algorithm has sharply decreased and increased again at 150 and 175 network size values is that due to the nature of the algorithm we used to construct the topology, more PLC nodes may be located in the destination area despite that the radius of the destination remains constant and the number of network nodes increases. In fact, because the topology construction algorithm does not expand much on the surface as the number of nodes increases, the network becomes more compact and new nodes are added in the empty holes between the nodes. The Hybrid algorithm, in fact, naturally faces more fluctuations in the cost of the tree and its other features due to the greedy choices it makes to select the node. But the Multiple algorithm has fewer fluctuations and is actually more stable because it involves all candidate nodes.

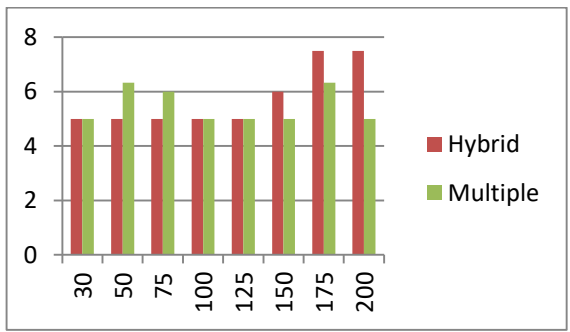

**Fig. 10**. The reliability of the proposed Hybrid and Multihop algorithms with respect to the number of nodes

Figure 10 does not provide much specific information about the behavior of the algorithms. The only important observation that can be made is that by increasing the network size, the Hybrid method achieves higher averages in reliability because it has more choices.

### 4.7. Maximum step parameter for PLC node to reach wireless node

Figure 11 shows the cost of the tree for different values of the maximum step for a PLC node to reach a wireless node. In this scenario, the radius of the destination area is constant and the value is 30 and the number of network nodes is 200. This graph has amazing results. The reason why the Hybrid method costs so much more for a step than the Multiple method is that when there is a step between PLC and wireless, there will usually be many candidate nodes in an area within a radius of 30. Only 2 of these candidate nodes are selected in the Hybrid method, which must cover all nodes in the area. But in the Multiple method, all the candidate nodes only cover the nodes around themselves. Thus, this condition creates a lower tree cost. As the maximum number of steps increases, the number of candidate nodes decreases, and consequently the tree cost of the Hybrid method decreases.

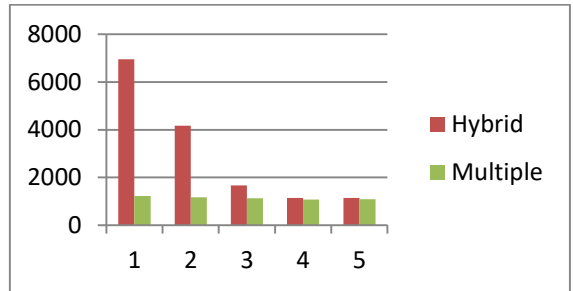

**Fig. 11**. The cost of the proposed Hybrid and Multihop algorithms with respect to the number of nodes

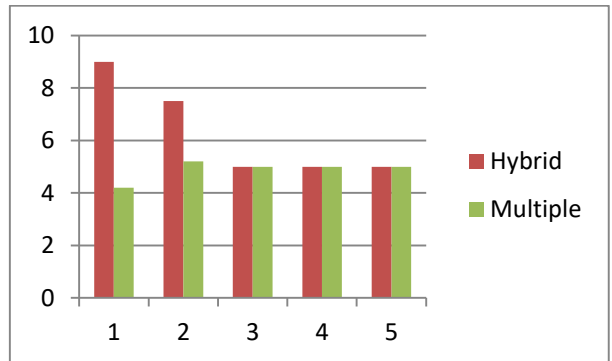

**Fig. 12.** The reliability of the proposed Hybrid and Multihop algorithms against the maximum step for reaching of PLC nodes

Figure 12 shows the average reliability of nodes against the maximum step for reaching of PLC nodes to wireless. As we explained in the graph, in maximum step 1, there are many choices for the Hybrid method, from which the two that have the highest reliability are selected. It leads to a higher average than the Multiple method. But as the maximum step increases, the two methods get closer to each other because the number of candidate nodes decreases and the Hybrid algorithm becomes more and more limited in its choices. Finally, one or two candidate nodes do not remain. In this case, there is a difference between these in the average reliability.

### 5.Conclusion

In this paper, two series of algorithms were presented to provide multicast security service in heterogeneous smart grids. The first series were algorithms worked with a single-step topology. That is, each PLC node had a wireless node next to itself with one step. MKMB and GCBT algorithms were proposed for this case. MKMB algorithm was an extension of KMB algorithm. The idea of GCBT method was to use geographical information in the CBT algorithm. An optimization model for constructing the Steiner tree was proposed and solved. The aim was to compare these algorithms with the optimal mode in terms of construction costs. These algorithms are compared with each other and evaluated for their performance. The result was that the MKMB algorithm is not much different from the usual KMB method in terms of performance, and the GCBT method tried to use geographic information alongside the CBT protocol, which its performance was good, but because it did not take into account the variable status of PLC connections, it cannot be implemented in smart grids. Also, two hybrid and multiple algorithms were presented. Their aim was to provide reliability for the nodes in the destination area by transmitting two wireless nodes into the destination area. This paper also compared the two algorithms in terms of tree cost and average reliability of nodes in different scenarios and interpreted the obtained results.

As the simulation results showed, the methods of the first part, ie single-step methods, were not very successful and they supported a certain topology, but at the same time, the efforts of this part led to the presentation of the algorithms of the second part, ie multi-step. Hybrid and Multiple methods are methods that try to increase the level of reliability for arriving packets by providing the idea of two trees in the destination area. They also provide 100% coverage for destination nodes. At the same time, they try to maximize the level of reliability and to reduce the delay of packets reaching the nodes. The simulation results confirm these claims.

Finally, the main achievements of the project are:

1. To provide reliability to ensure that packets are received by nodes of the destination area by creating two optimal trees to transmit packets to each node.
2. To consider the security issues by selecting the wireless nodes with the most reliability.
3. 100% coverage of nodes in the destination area in terms of receiving multicast packets.
4. To reduce delay for arriving packets to destination nodes, which is useful for delay-sensitive applications in smart grids.
5. To provide a way to incorporate geographic information into the KMB algorithm in order to use this additional information for better routing in packet multicast process.
6. To incorporate geographic information into the CBT algorithm and provide GCBT method

**REFERENCES**

[1]. Thomas, Y., Fotiou, N., Toumpis, S., & Polyzos, G. C. (2020). Improving mobile ad hoc networks using hybrid IP-information centric networking. Computer Communications, 156, 25-34.

[2]. Chen, J., Yan, F., Li, D., Chen, S., & Qiu, X. (2020). Recovery and Reconstruction of Multicast Tree in Software-Defined Network: High Speed and Low Cost. IEEE Access, 8, 27188-27201.

[3]. Bijur, G., Ramakrishna, M., & Kotegar, K. A. (2021). Multicast tree construction algorithm for dynamic traffic on software defined networks. Scientific Reports, 11(1), 1-15.

[4]. Li, B., & Wang, J. (2021). An identifier and locator decoupled multicast approach (ILDM) based on ICN. Applied Sciences, 11(2), 578.

[5]. Liu, Q., Tang, R., Ren, H., & Pei, Y. (2020). Optimizing multicast routing tree on application layer via an encoding-free non-dominated sorting genetic algorithm. Applied Intelligence, 50(3), 759-777.

[6]. Godinez, F., Tomi-Tricot, R., Quesson, B., Barthel, M., Lykowsky, G., Scott, G., ... & Malik, S. (2021). An 8 channel parallel transmit system with current sensor feedback for MRI-guided interventional applications. Physics in Medicine & Biology, 66(21), 21NT05.

[7]. Deldouzi, S., & Coutinho, R. W. (2021, November). A Novel Harvesting-Aware RL-based Opportunistic Routing Protocol for Underwater Sensor Networks. In Proceedings of the 24th International ACM Conference on Modeling, Analysis and Simulation of Wireless and Mobile Systems (pp. 87-94).

[8]. Jahandideh, Y., & Mirzaei, A. (2021). Allocating Duplicate Copies for IoT Data in Cloud Computing based on Harmony Search Algorithm. IETE Journal of Research, 1-14..

[9]. Javid, S., & Mirzaei, A. (2021). Presenting a Reliable Routing Approach in IoT Healthcare Using the Multiobjective-Based Multiagent Approach. Wireless Communications and Mobile Computing, 2021.

[10] Bavaghar, M., Mohajer, A., & Taghavi Motlagh, S. (2020). Energy Efficient Clustering Algorithm for Wireless Sensor Networks. Journal of Information Systems and Telecommunication (JIST), 4(28), 238.

[11] Rahimi, A. M., Ziaeddini, A., & Gonglee, S. (2021). A novel approach to efficient resource allocation in load-balanced cellular networks using hierarchical DRL. Journal of Ambient Intelligence and Humanized Computing, 1-15.

[12] Zhang, S., Madadkhani, M., Shafieezadeh, M., & Mirzaei, A. (2019). A novel approach to optimize power consumption in orchard WSN: Efficient opportunistic routing. Wireless Personal Communications, 108(3), 1611-1634.

[13] Mirzaei, A., Barari, M., & Zarrabi, H. (2019). Efficient resource management for non-orthogonal multiple access: A novel approach towards green hetnets. Intelligent Data Analysis, 23(2), 425-447.

[14] Mirzaei, A., Zandiyan, S., & Ziaeddini, A. (2021). Cooperative Virtual Connectivity Control in Uplink Small Cell Network: Towards Optimal Resource Allocation. Wireless Personal Communications, 1-25.

[15] Nikjoo, F., Mirzaei, A., & Mohajer, A. (2018). A novel approach to efficient resource allocation in NOMA heterogeneous networks: Multi-criteria green resource

management. Applied Artificial Intelligence, 32(7-8), 583-612.

[16] Somarin, A. M., Alaei, Y., Tahernezhad, M. R., Mohajer, A., & Barari, M. (2015). An Efficient Routing Protocol for Discovering the Optimum Path in Mobile Ad Hoc Networks. Indian Journal of Science and Technology, 8(S8), 450-455.

[17] Mohajer, A., Yousefvand, M., Ghalenoo, E. N., Mirzaei, P., & Zamani, A. (2014). Novel approach to sub-graph selection over coded wireless networks with QoS constraints. IETE Journal of Research, 60(3), 203-210.

[18] Mohajer, A., Bavaghar, M., Saboor, R., & Payandeh, A. (2013, August). Secure dominating set-based routing protocol in MANET: Using reputation. In 2013 10th International ISC Conference on Information Security and Cryptology (ISCISC) (pp. 1-7). IEEE.

[19] Mirzaei, A., Barari, M., & Zarrabi, H. (2021). An Optimal Load Balanced Resource Allocation Scheme for Heterogeneous Wireless Networks based on Big Data Technology. arXiv preprint arXiv:2101.02666.

[20] Farhang, M., Mohajer, A., Zobeyravi, O., & Rahimzadegan, A. Adaptive Spectrum Sensing Algorithm Based on Noise Variance Estimation for Cognitive Radio Applications.

[21] Mohajer, A., Somarin, A., Yaghoobzadeh, M., & Gudakahriz, S. (2016). A Method Based on Data Mining for Detection of Intrusion in Distributed Databases. Journal Of Engineering And Applied Sciences, 11(7), 1493-1501.

[22] Mohajer, A., Hajimobini, M. H., Mirzaei, A., & Noori, E. (2014). Trusted-CDS Based Intrusion Detection System in Wireless Sensor Network (TC-IDS). Open Access Library Journal, 1(7), 1-10.

[23]. Mirzaei, A., & Najafi Souha, A. (2021). Towards optimal configuration in MEC Neural networks: deep learning-based optimal resource allocation. Wireless Personal Communications, 121(1), 221-243.

[24]. Mohajer, A., Barari, M., & Zarrabi, H. (2016). Big data-based self optimization networking in multi carrier mobile networks. Bulletin de la Société Royale des Sciences de Liège, 85, 392-408.

[25]. Mohajer, A., Mazoochi, M., Niasar, F. A., Ghadikolayi, A. A., & Nabipour, M. (2013, June). Network Coding-Based QoS and Security for Dynamic Interference-Limited Networks. In International Conference on Computer Networks (pp. 277-289). Springer, Berlin, Heidelberg.

[26]. Mohajer, A., Bavaghar, M., & Farrokhi, H. (2020). Reliability and mobility load balancing in next generation self-organized networks: using stochastic learning automata. Wireless Personal Communications, 114(3), 2389-2415.

[27]. Xiao, J., Liu, Y., Qin, H., Li, C., & Zhou, J. (2021). A Novel QoS Routing Energy Consumption Optimization Method Based on Clone Adaptive Whale Optimization Algorithm in IWSNs. Journal of Sensors, 2021.

[28]. Djamaa, B., Senouci, M. R., Bessas, H., Dahmane, B., & Mellouk, A. (2021). Efficient and stateless P2P routing mechanisms for the Internet of Things. *IEEE Internet of Things Journal*, *8*(14), 11400-11414.

[29]. Xiao, Z., Li, Y., & Zhou, H. (2021, April). Research on Remote Online Teaching Assistant System Based on Human-Computer Interaction. In International Conference on Multimedia Technology and Enhanced Learning (pp. 203-215). Springer, Cham.

[30]. Xiao, Z., Li, Y., & Zhou, H. (2021, April). Research on Remote Online Teaching Assistant System Based on Human-Computer Interaction. In International Conference on Multimedia Technology and Enhanced Learning (pp. 203-215). Springer, Cham.

[31]. Yang, F., Han, J., Ding, X., & Zhao, C. (2021). Optimal resource utilization for intra-cluster D2D retransmission and cooperative communications in VANETs. Wireless Networks, 27(5), 3251-3271.

[32]. Escobar-Molero, A., Heinen, S., & Mähönen, P. (2020). Using concurrent transmissions to improve the reliability and latency of low-power wireless mesh networks (No. RWTH-2020-05541). Lehrstuhl für Integrierte Analogschaltungen und Institut für Halbleitertechnik.

[33]. Mirzaei Somarin, A., Barari, M., & Zarrabi, H. (2018). Big data based self-optimization networking in next generation mobile networks. Wireless Personal Communications, 101(3), 1499-1518.

[34]. Xiao, C., Lou, H., Li, C., & Jin, K. (2020, August). DBM: A dimension-bubble-based multicast routing algorithm for 2D mesh network-on-chips. In Conference on Advanced Computer Architecture (pp. 43-55). Springer, Singapore.

[35]. Li, X., Lan, X., Mirzaei, A., Aghdam, M. J., & Member, I. E. E. E. (2021). Reliability and Robust Resource Allocation for Cache-enabled HetNets: QoS-aware Mobile Edge Computing. Reliability Engineering & System Safety, 108272.